\documentclass[aps,prc,twocolumn,superscriptaddress]{revtex4-2}
\usepackage{xcolor}
\usepackage{graphicx}
\usepackage{subfig}    
\usepackage{overpic}
\usepackage{caption}
\usepackage{booktabs} 
\usepackage{enumitem}

\begin{document}


\title{Particle productions in $p\bar{p}$ collisions in the PACIAE 4.0 model}


\author{Zhen Xie}
\affiliation{School of Physics and Information Technology, Shaanxi Normal University, Xi'an 710119, China}


\author{An-Ke Lei}
\affiliation{School of Physics and Electronic Science, Guizhou Normal University, Guiyang, 550025, China}

\author{Hua Zheng}
\email[]{zhengh@snnu.edu.cn}

\author{Wenchao Zhang}
\affiliation{School of Physics and Information Technology, Shaanxi Normal University, Xi'an 710119, China}

\author{Dai-Mei Zhou}
\email[]{zhoudm@mail.ccnu.edu.cn}
\affiliation{Key Laboratory of Quark and Lepton Physics (MOE) and Institute of Particle Physics, Central China Normal University, Wuhan 430079, China}

\author{Zhi-Lei She}
\affiliation{School of Mathematical and Physical Sciences, Wuhan Textile University, Wuhan 430200, China}

\author{Yu-Liang Yan}
\affiliation{China Institute of Atomic Energy, P. O. Box 275 (10), Beijing 102413, China}

\author{Ben-Hao Sa}
\email[]{sabhliuym35@qq.com}
\affiliation{Key Laboratory of Quark and Lepton Physics (MOE) and Institute of Particle Physics, Central China Normal University, Wuhan 430079, China}
\affiliation{China Institute of Atomic Energy, P. O. Box 275 (10), Beijing 102413, China}


\date{\today}

\begin{abstract}
We investigate the particle production in proton–antiproton ($p\bar{p}$) collisions using the PACIAE 4.0 model. The pseudorapidity density distributions ($dN_{\text{ch}}/d\eta$) and transverse momentum ($p_T$) spectra of charged particles from nonsingle diffractive (NSD) $p\bar{p}$ collisions agree well with the experimental data when using model parameters previously determined from nonsingle diffractive proton-proton ($pp$) collisions. Furthermore, we systematically compare results from both inelastic (INEL) and nonsingle diffractive $p\bar{p}$ and $pp$ collisions at the same energy to study the effect of the initial state (matter vs. antimatter) on the transverse momentum spectra of identified particles. Our results show that the net baryon-number difference in the initial state significantly enhances nucleon production at low collision energies, while its effect becomes negligible for high-multiplicity particles or at high collision energies, as expected. These findings further prove that the PACIAE 4.0 model is a versatile and reliable tool for studying high-energy collision physics.
\end{abstract}


\maketitle

\section{INTRODUCTION}{\label{intro}}
The pseudorapidity density distributions of charged particles ($dN_{ch}/d\eta$) and the transverse momentum ($p_T$) spectra of charged (identified) particles are fundamental observables for understanding the physics in high-energy elementary particle and heavy-ion collisions~\cite{PHOBOS:2010eyu,STAR:2006xud,STAR:2006nmo,ALICE:2013jfw,ALICE:2015olq,ALICE:2019hno,ALICE:2014juv,ALICE:2015ial,ALICE:2015qqj,ALICE:2010cin,ALICE:2011gmo,ALICE:2022kol}.  These distributions primarily reflect the dynamics of final-state interactions and the properties of the system at freeze-out, while also retaining valuable information about the initial collision dynamics and early-stage evolution. Experimental collaborations have collected and analyzed extensive data on these observables to study properties of the quark-gluon plasma (QGP), the particle physics as well as to search for new physics beyond the Standard Model (SM) ~\cite{PHOBOS:2010eyu,STAR:2006xud,STAR:2006nmo,ALICE:2013jfw,ALICE:2015olq,ALICE:2019hno,ALICE:2014juv,ALICE:2015ial,ALICE:2015qqj,ALICE:2010cin,ALICE:2011gmo,ALICE:2022kol}. In parallel, theorists have developed various models, such as PYTHIA~\cite{Sjostrand:2006za,Bierlich:2022pfr},  HERWIG~\cite{Corcella:2000bw}, HIJING~\cite{Wang:1991hta}, UrQMD~\cite{Bass:1998ca,Bleicher:1999xi}, AMPT~\cite{Lin:2004en}, PACIAE~\cite{Sa:2011ye,Lei:2023srp,Lei:2024kam}, EPOS-LHC~\cite{PhysRevC.92.034906}, SMASH~\cite{SMASH:2016zqf}, and Hydrodynamics ~\cite{Song:2010mg,Gale:2013da} to simulate and study the physics of high-energy collisions~\cite{Xie:2025vnh,Xie:2025zoi,Zhang:2024kdv,Werner:2024ntd,Yang:2022fcj,Zhang:2025pqu,STAR:2017sal,Toia:2011nzq,Deppman:2019yno,Shi:2024pyz,Tao:2023kcu,Tao:2022tcw,Zhu:2022dlc,Zhu:2022bpe,Tao:2020uzw,Zhu:2021fbs,Zheng:2015mhz,Wang:2022det,Zhu:2018nev,Gao:2017yas,Zheng:2015gaa,Zheng:2015tua,Wong:2015mba,Rath:2019cpe,Xu:2017akx,Yang:2022fcj,Zhao:2020wcd,Zhao:2021vmu,Tan:2024lrp}. To refine and validate these models, their simulated results, obtained with tuned parameters, are compared with the experimental data, ideally as many observables as possible. 

Both proton-antiproton ($p\bar{p}$) and proton-proton $(pp)$ collisions are elementary processes in high-energy physics, providing an essential baseline for understanding the physics of heavy-ion collisions. Given their importance, these elementary systems have been studied extensively in experiments. The available experimental data enables theorists to conduct systematic simulations and constrain their model parameters. Conversely, with refined models, theorists and experimentalists can gain a deeper understanding of the underlying physics and provide reliable simulated data for regions where experimental results are not yet available~\cite{Xie:2025vnh, Xie:2025zoi} and references therein.

PACIAE is a multipurpose Monte Carlo event generator based on the PYTHIA model. It is designed to simulate a wide range of high-energy collisions, including lepton-lepton, lepton-hadron, lepton-nucleus,
hadron-hadron, hadron-nucleus, and nucleus-nucleus collisions. In our previous works \cite{Xie:2025vnh,Xie:2025zoi}, we conducted the systematic investigations of the inelastic (INEL) $pp$ collisions and the nonsingle diffracitve (NSD) $pp$ collisions using the latest version, PACIAE 4.0,  across a wide range of energies from 200 GeV to the top LHC energy. The model, with tuned parameters, successfully reproduces the pseudorapidity density distributions of charged particles and the transverse momentum spectra of identified (charged) particles. 

Compared to proton-proton ($pp$) collisions, a distinguishing feature of proton-antiproton ($p\bar{p}$) collisions is the potential for annihilation between the proton and antiproton. This property made $p\bar{p}$ collisions a historically important venue for providing clean access to electroweak processes, such as the production of $W^{\pm}/Z$ bosons~\cite{D0:1997hsd,CDF:1998tqc,Warburton:2004vd}. Furthermore, by comparing observables from $pp$ and $p\bar{p}$ collisions under identical event selections and collision energies, one can gain unique insights into the differences between matter-antimatter and mater-matter interactions. Similar to the idea of probing the interaction between antiprotons in Au+Au collisions conducted by the STAR collaboration at the
Relativistic Heavy Ion Collider (RHIC) \cite{STAR:2015kha}. 

The UA1, CDF, and P238 collaborations have published data on the pseudorapidity density distributions and transverse momentum spectra of charged particles produced in NSD $p\bar{p}$ collisions at center-of-mass energies of $\sqrt{s}=0.54$, 0.63, and 1.8  TeV~\cite{UA1:1982yyh,Harr:1997sa,CDF:1989nkn,UA1:1982fux,CDF:1988evs}. In this work, we test the robustness of the PACIAE 4.0 model by applying the parameter set previously determined for NSD $pp$ collisions directly to NSD $p\bar{p}$ collisions, without any further adjustment. We simulate the pseudorapidity density distributions and transverse momentum spectra of charged particles and compare them with the corresponding experimental data. Furthermore, to investigate the influence of matter-antimatter interactions on particle production, we compare the transverse momentum spectra of identified particles from both inelastic and nonsingle diffractive $pp$ and $p\bar{p}$ collisions at various energies. 

The paper is organized as follows. Section II provides a brief introduction to the PACIAE 4.0 model. In Sec. III, we present the simulation results for pseudorapidity density distributions and transverse momentum spectra of charged particles in NSD $p\bar{p}$ collisions at $\sqrt{s} = 0.54$, 0.63, and 1.8 TeV, including comparisons with experimental data. The influence of matter-antimatter interactions on particle production by comparing the transverse momentum spectra of identified particles from both inelastic and nonsingle diffractive $pp$ and $p\bar{p}$ collisions at given energies are also discussed. A summary is given in Sec. IV.

\begin{figure}[h]
\centering
\includegraphics[width=1\linewidth]{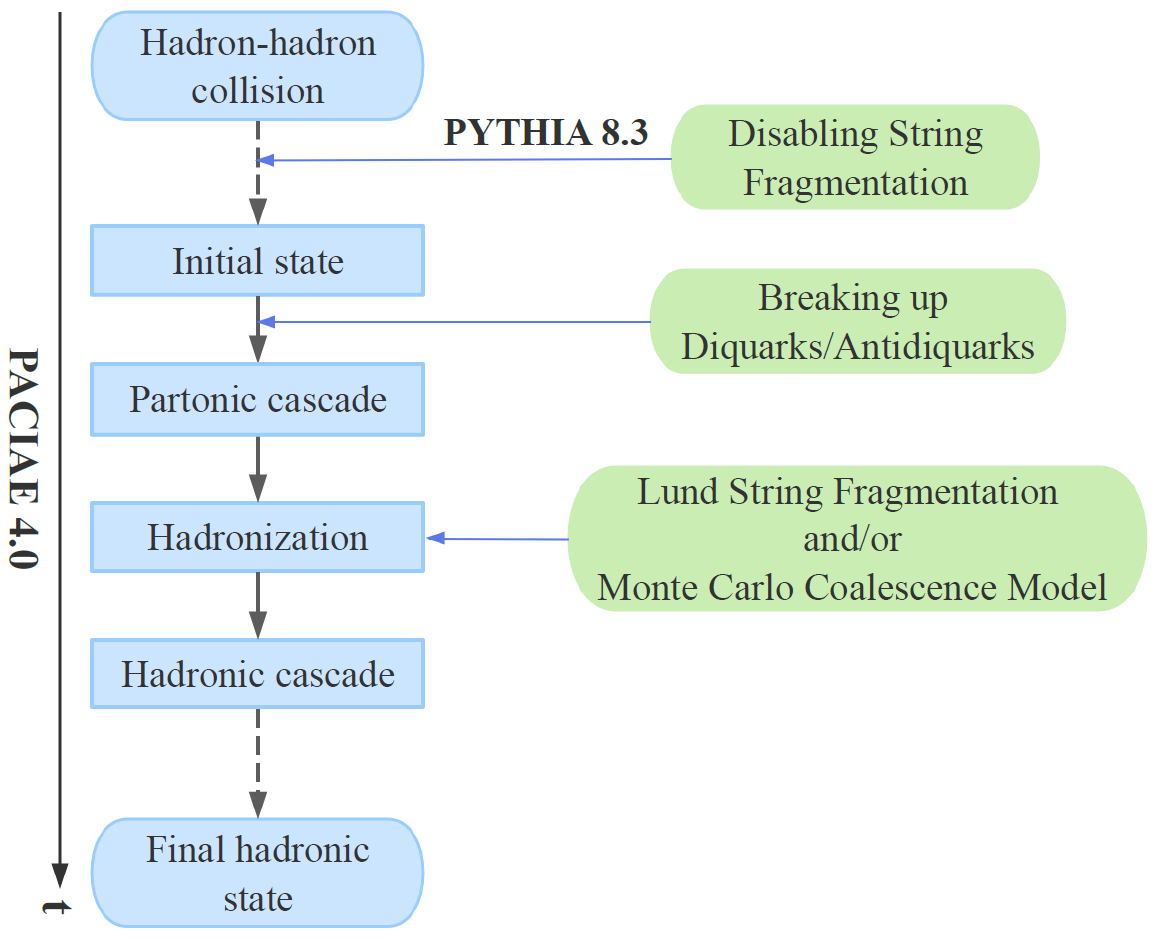}
\caption{The program flow for a $pp(\bar{p})$ collision in the PACIAE 4.0 model.}
\label{fig:flow}
\end{figure}

\section{Brief introduction to PACIAE 4.0 model}\label{mod}
The PACIAE 4.0 model~\cite{Sa:2011ye,Lei:2023srp,Lei:2024kam} is a Monte Carlo event generator based on PYTHIA 8.3, incorporating partonic rescattering (PRS) in the partonic phase and the hadronic rescattering (HRS) after hadronization. The model consists of four stages, outlined as follows:
\begin{enumerate}[label=(\roman*)]
\item Initial state: The initial partonic state is generated by executing PYTHIA 8.3~\cite{Sjostrand:2006za, Bierlich:2022pfr} with the Lund string fragmentation temporary disabled and postsetting of breaking up diquarks/antidiquarks randomly, along with the initial- and final-state radiation.  
\item Partonic rescattering: Partonic rescattering is modeled through $2 \rightarrow2$ parton-parton scattering processes using leading-order perturbative quantum chromadynamics (pQCD) cross sections. The differential cross section is given by
\begin{equation}
    \frac{d\sigma}{dt}(ab\rightarrow cd;s,t)= K\frac{\pi\alpha_s^2}{s^2}|\bar{M}(ab\rightarrow cd)|^2,
\end{equation}
where $\alpha_s$ is the strong coupling constant. The variables $s, t$ are the Mandelstam invariants in the kinematics of the $ab\rightarrow cd$ process. The parameter $K$ is a multiplicative correction factor for the hard scattering cross sections. We refer to Ref. \cite{Lei:2023srp} for the details of $|\bar{M}(ab\rightarrow cd)|^2$.
\item Hadronization: The above obtained final partonic state is hadronized using the Lund string fragmentation mechanism in this work. The model also supports another hadronization mechanism, i.e., coalescence. The Lund string fragmentation function~\cite{Sjostrand:2006za,Andersson:1983ia} is
\begin{equation}
    f(z) \propto (1/z) (1-z)^a \exp(-b m_T^2 / z),
\end{equation}
where the parameters $a$ and $b$ are pivotal for particle production. Here, $z$ is the energy fraction carried away by the produced hadron, and $m_T=\sqrt{m^2+p_T^2}$ is its transverse mass.
\item Hadronic rescattering: PACIAE 4.0 provides nearly 600 different inelastic hadron-hadron ($hh$) collision channels in addition to elastic $hh$ collisions. The resulted hadrons undergo $2\rightarrow2$ rescatterings with experimentally and/or empirically determined cross sections until kinetic freeze-out, producing the final-state particles. 
\end{enumerate}
Figure ~\ref{fig:flow} illustrates the program flow for a $pp(\bar{p})$ collision in PACIAE 4.0. The effects of partonic and hadronic rescatterings on particle yields are significant in high-energy particle collisions and are discussed in detail in Ref. \cite{Xie:2025vnh}. In this work, PACIAE 4.0 is run with both partonic and hadronic rescatterings enabled. 
\begin{figure*}[!ht]
   \centering
    \subfloat{
	\begin{overpic}[width=0.48\linewidth]{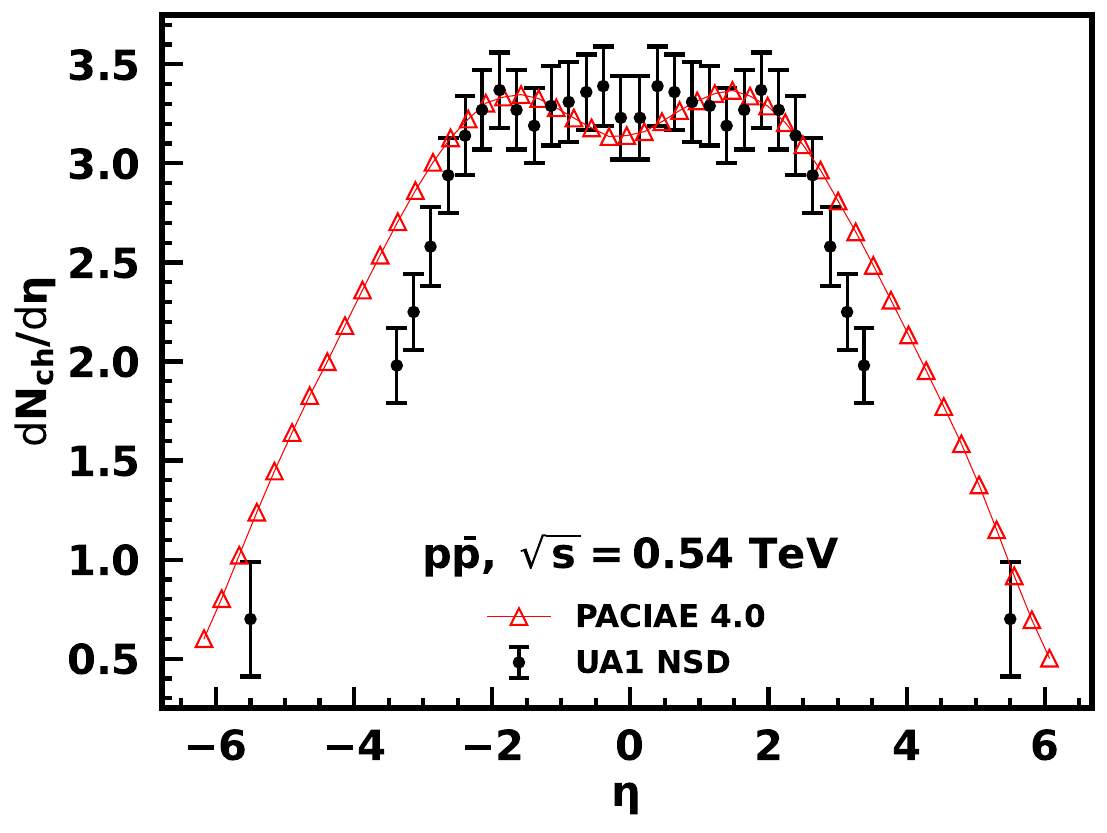}
    \end{overpic}
	}
    \hfill
     \subfloat{
	\begin{overpic}[width=0.48\linewidth]{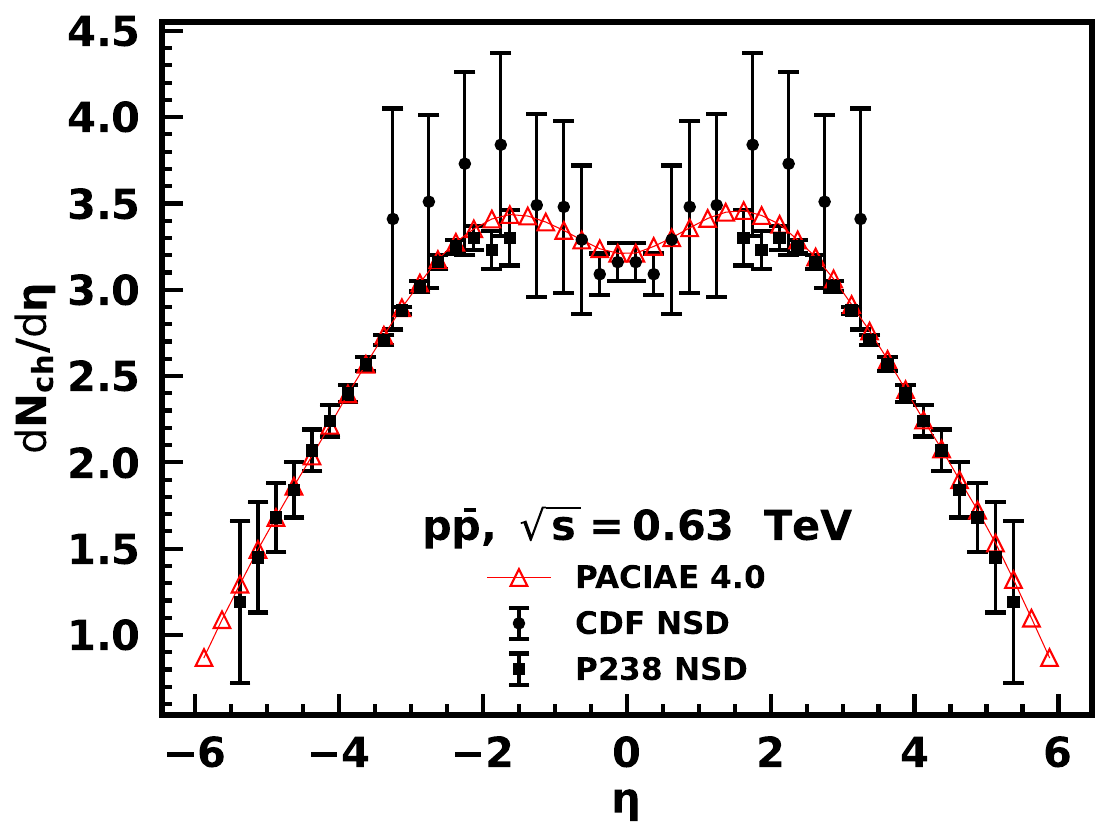}
    \end{overpic}
	}
    \hfill
    \subfloat{
	\begin{overpic}[width=0.48\linewidth]{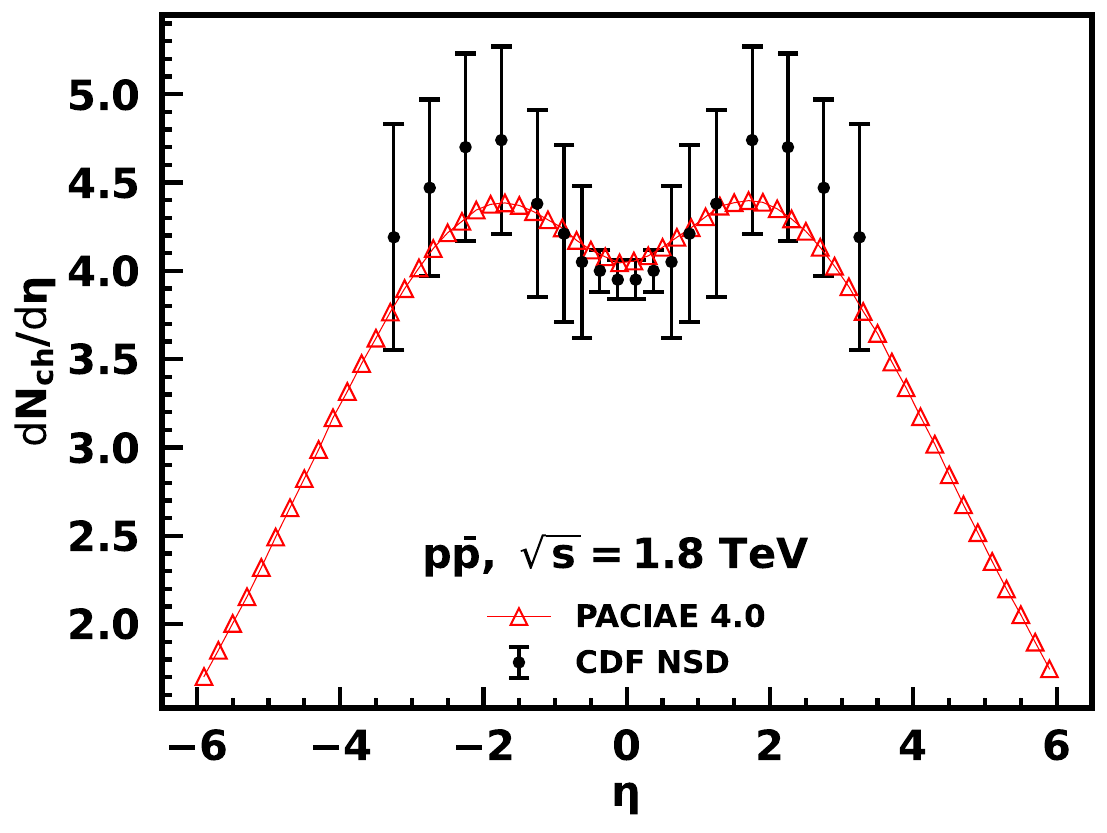}
    \end{overpic}
	}
    \hfill
   \caption{The pseudorapidity density distributions of charged particles produced in NSD $p\bar{p}$ collisions at $\sqrt{s}= 0.54, 0.63, 1.8$ TeV from PACIAE 4.0 model simulations compared with experimental data from UA1, CDF and P238 collaborations \cite{UA1:1982yyh,Harr:1997sa,CDF:1989nkn}.}
  \label{fig1}
\end{figure*}

\begin{figure}[!ht]
   \centering
    \subfloat{
	\begin{overpic}[width=1.0\linewidth]{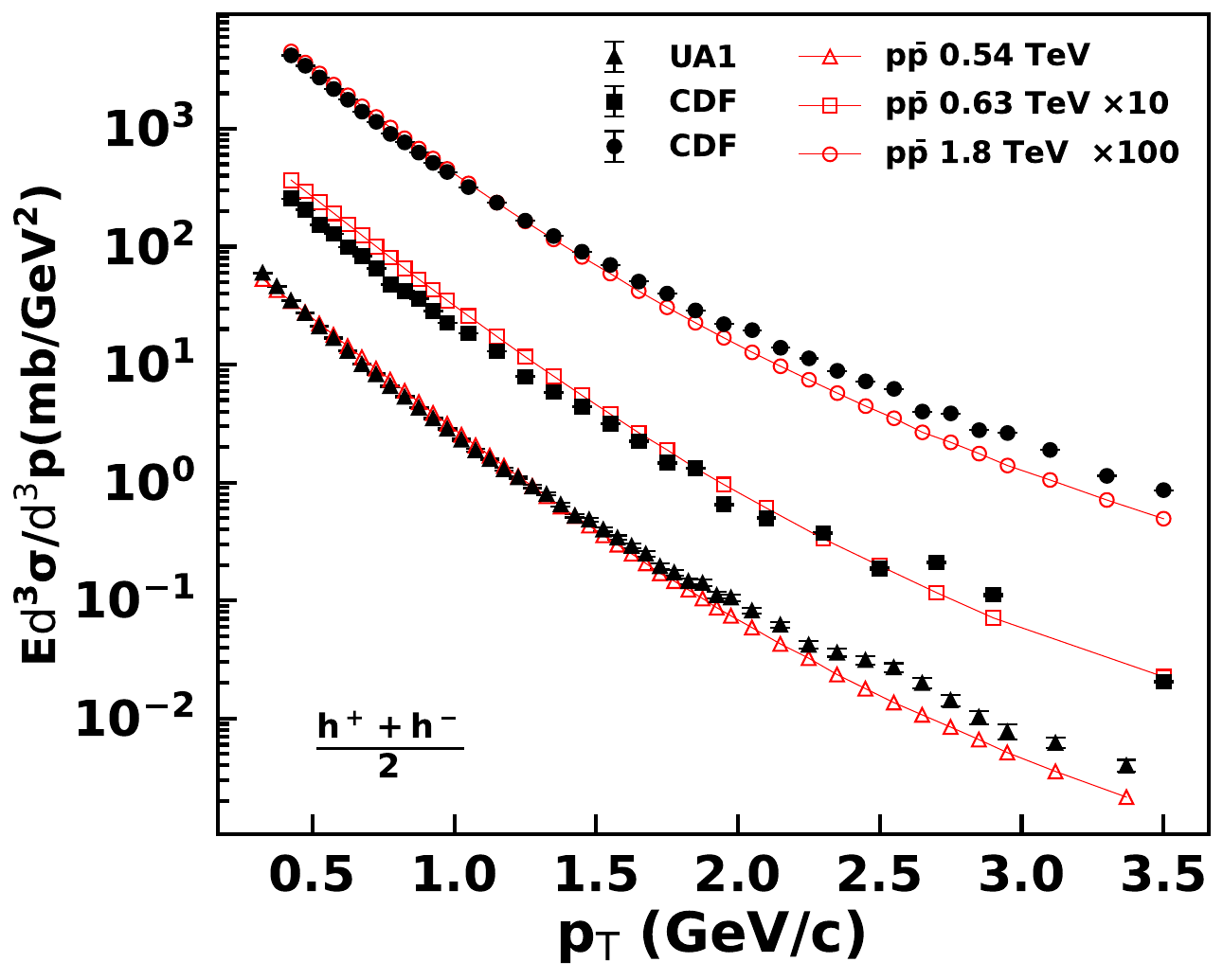}
    \end{overpic}
	}
    \hfill
   \caption{The transverse momentum spectra of charged particles produced in NSD $p\bar{p}$ collisions at $\sqrt{s}= 0.54, 0.63, 1.8$ TeV from PACIAE 4.0 model simulations compared with UA1 and CDF experimental data \cite{UA1:1982fux,CDF:1988evs}.}
  \label{fig2-1}
\end{figure}

\begin{figure*}[!ht]
   \centering
    \subfloat{
	\begin{overpic}[width=0.485\linewidth]{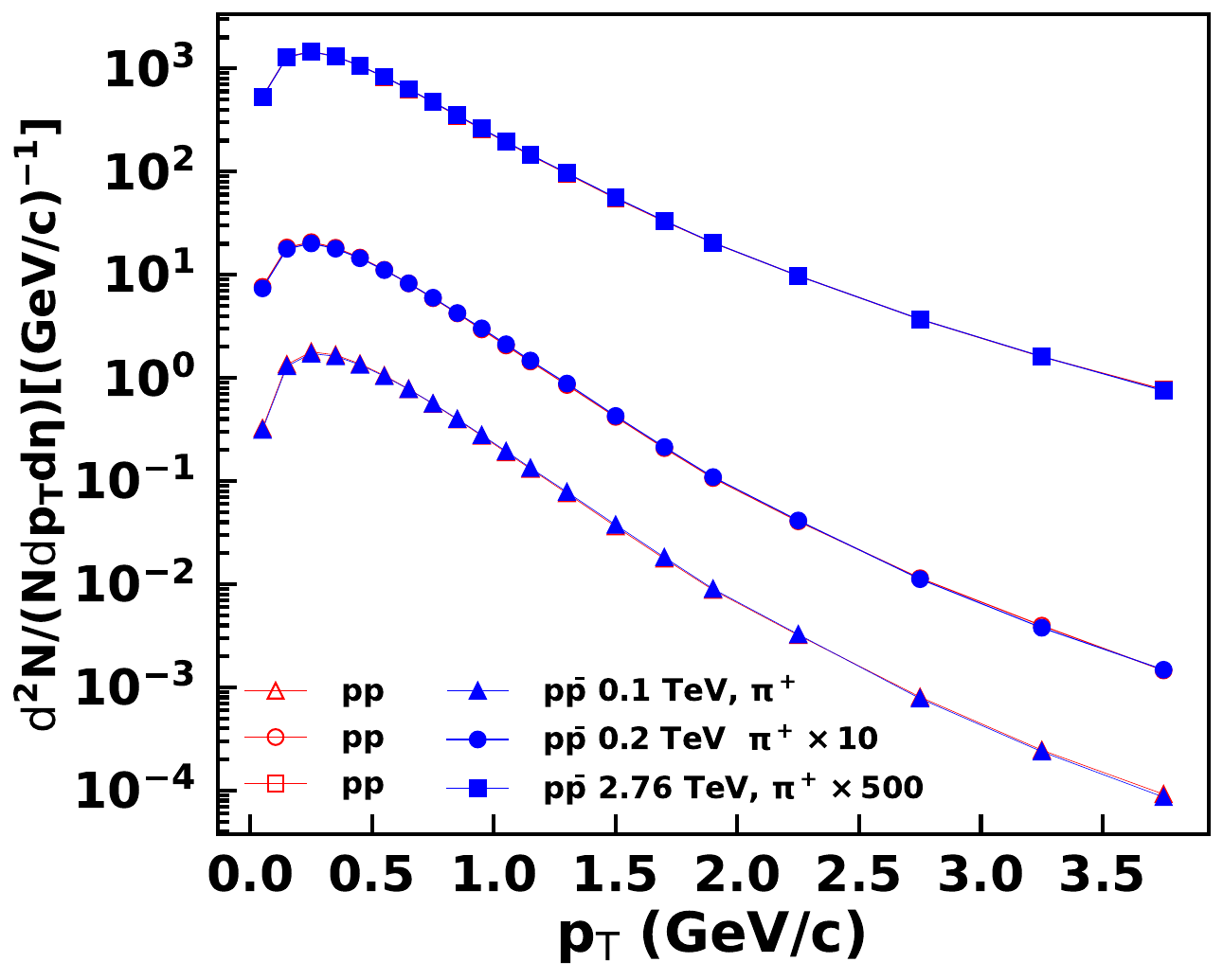}
    \end{overpic}
	}
    \hfill
     \subfloat{
	\begin{overpic}[width=0.485\linewidth]{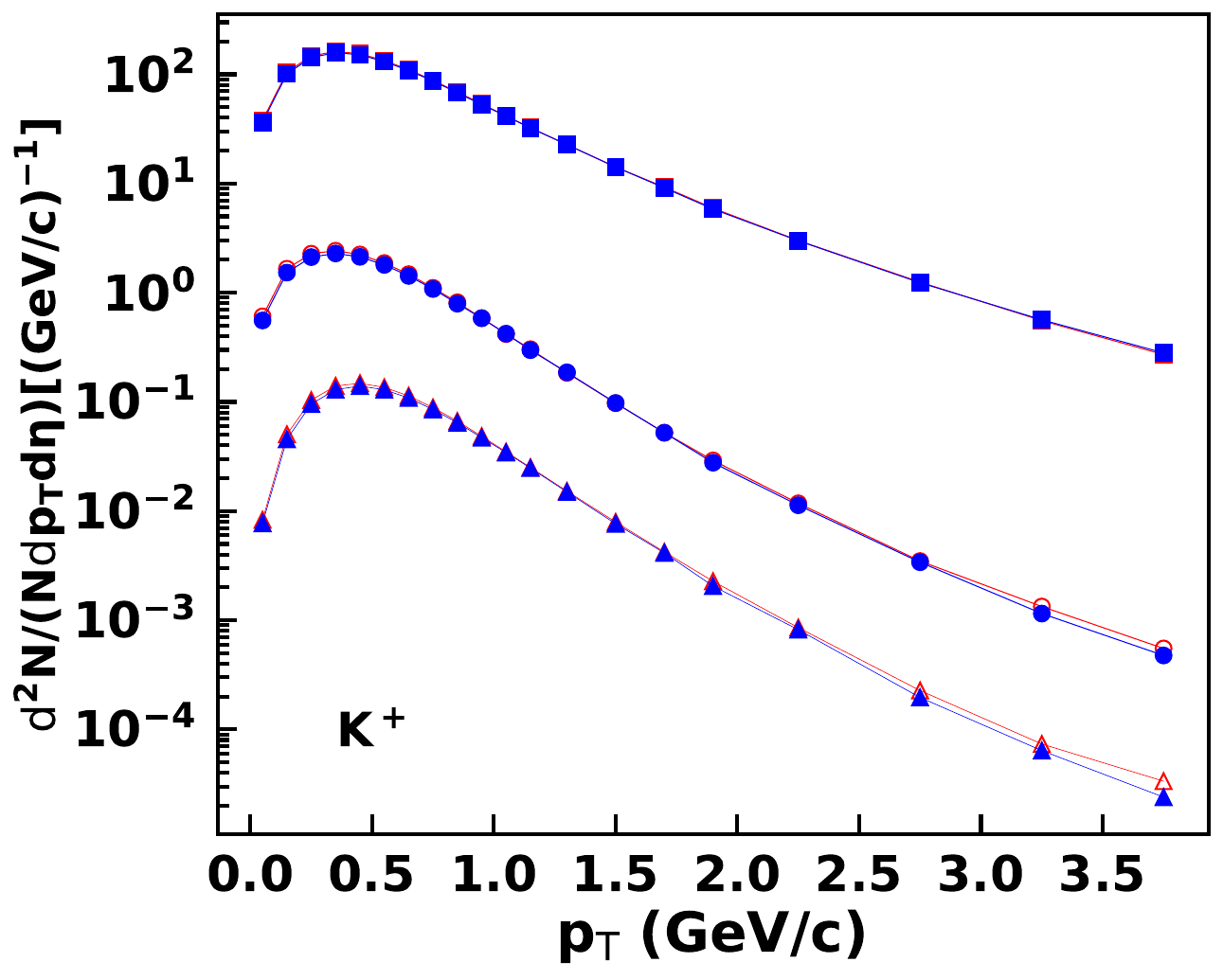}
    \end{overpic}
	}
    \hfill
    \subfloat{
	\begin{overpic}[width=0.485\linewidth]{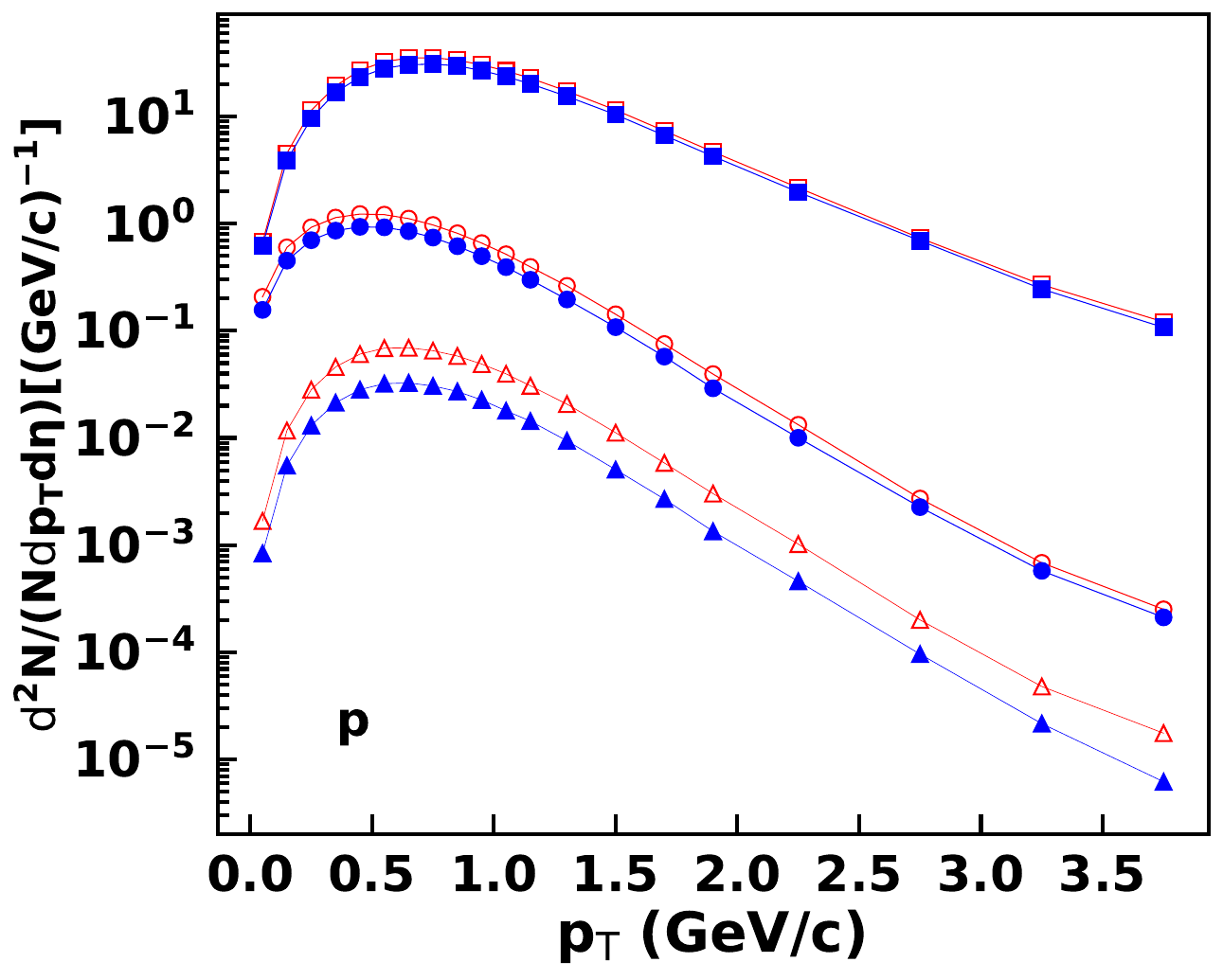}
    \end{overpic}
	}
    \hfill
    \subfloat{
	\begin{overpic}[width=0.485\linewidth]{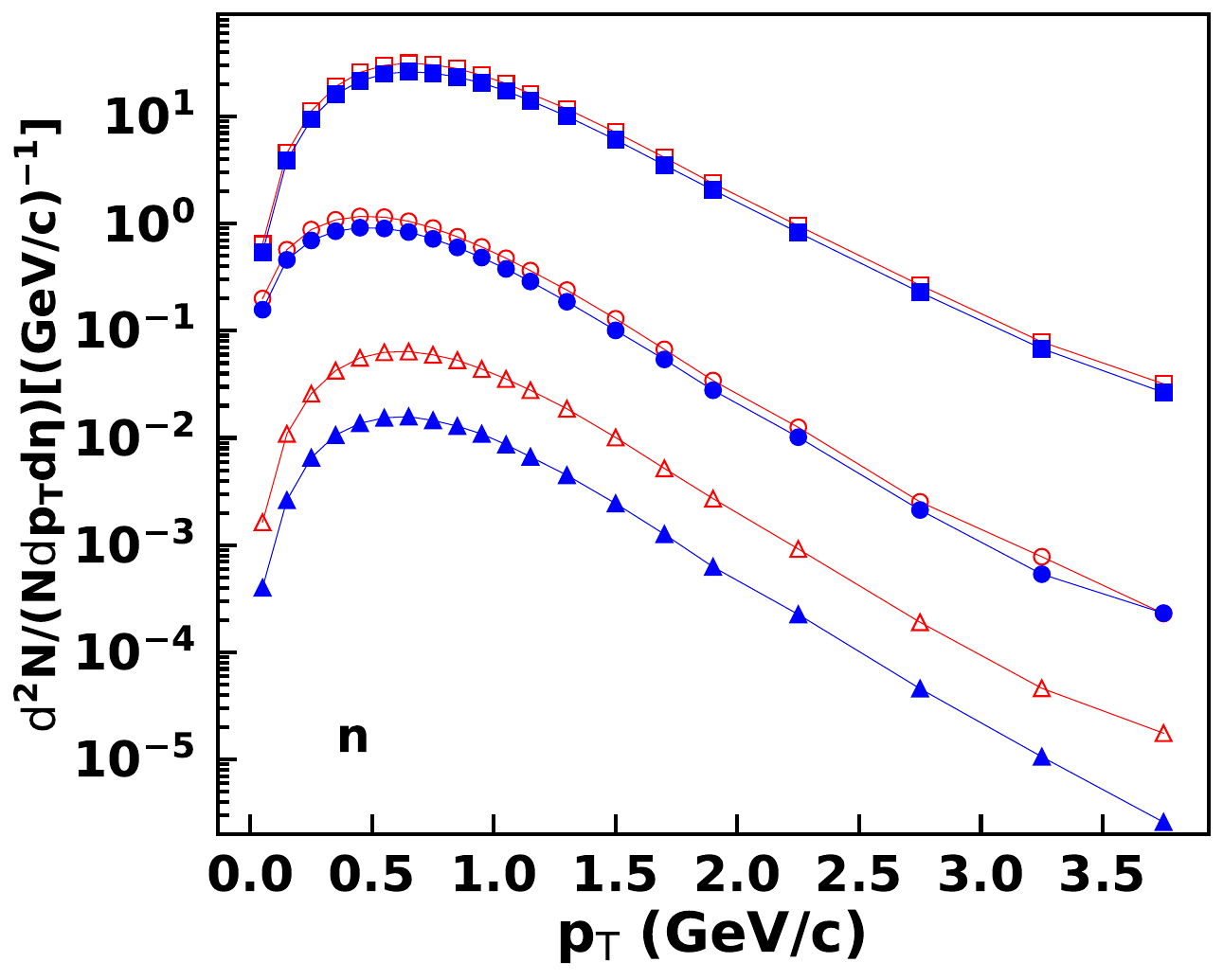}
    \end{overpic}
	}
    \hfill
   \caption{The transverse momentum spectra of $\pi^+$, $K^+$, protons and neutrons produced in INEL $p\bar{p}$ collisions and $pp$ collisions at $\sqrt{s}= 0.1, 0.2, 2.76$ TeV from PACIAE 4.0 simulations.}
  \label{fig2}
\end{figure*}

\begin{figure}[!ht]
   \centering
    \subfloat{
	\begin{overpic}[width=1.0\linewidth]{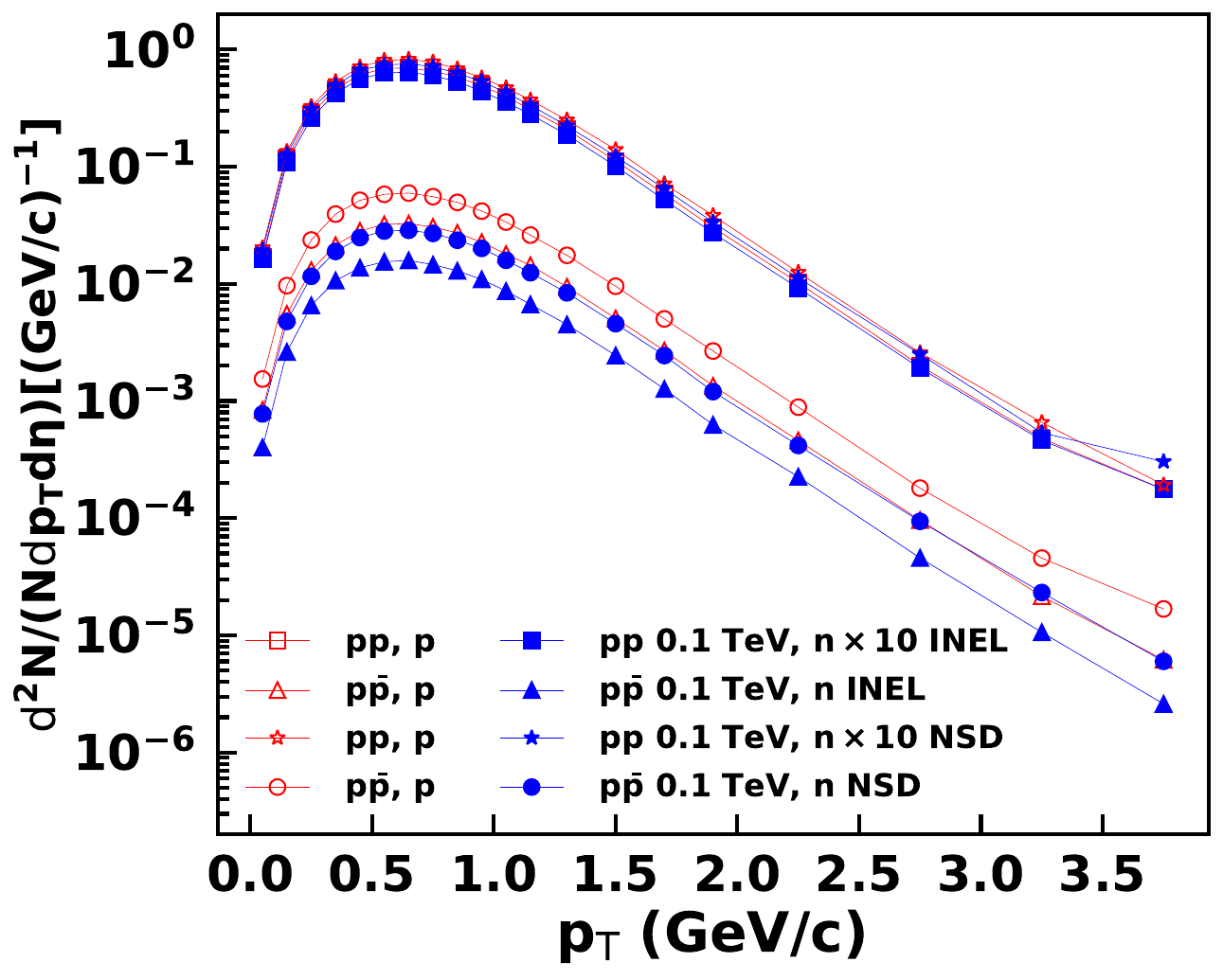}
    \end{overpic}
	}
   \caption{Comparison of transverse momentum spectra of protons and neutrons produced in INEL(NSD) $p\bar{p}$ collisions and $pp$ collisions at $\sqrt{s}= 0.1$ TeV from PACIAE 4.0 simulations. The data for INEL collisions are the same as those in Fig. \ref{fig2}.}
  \label{fig3}
\end{figure}

\section{RESULTS AND DISCUSSION }\label{rad}
Using the PACIAE 4.0 model parameters previously determined for INEL and NSD $pp$ collisions~\cite{Xie:2025vnh,Xie:2025zoi}, and assuming these parameters are also valid for the same event types in $p\bar{p}$ collisions, we simulate the pseudorapidity density distributions of charged particles and the transverse momentum spectra of charged (identified) particles. The key parameters are listed below: \\

\begin{enumerate}[label=(\arabic*)]
    \item The parameter $K$ is fixed at 0.8 and the parameter $a$ is fixed at 0.68 for both NSD and INEL $p\bar{p}$ collision events, consistent with their values in $pp$ collisions ~\cite{Xie:2025zoi,Xie:2025vnh}.
    \item For NSD events, the parameter $b$ is fixed at 0.8 for all investigated collision energies in $p\bar{p}$ collisions, the same as for NSD events in $pp$ collisions~\cite{Xie:2025zoi}. For INEL events, $b$ is calculated from the empirical function $0.646 erf(\frac{1}{2}\ln\sqrt{s}-4.37)+1.374$ which was determined from a systematic study of the inelatic $pp$ collisions, where $\sqrt{s}$ is in units of GeV~\cite{Xie:2025vnh}. 
\end{enumerate}
The other input parameters are set to their default values in PACIAE 4.0. We emphasize that all results for $p\bar{p}$ collisions presented in the following are simulated using PACIAE 4.0 without any adjustable model parameters.

In Fig.~\ref{fig1}, we show the pseudorapidity density distributions of charged particles produced in NSD $p\bar{p}$ collisions at $\sqrt{s}=$ 0.54, 0.63, and 1.8 TeV. The experimental data from the UA1, CDF, and P238 collaborations~\cite{UA1:1982yyh,Harr:1997sa,CDF:1989nkn} are represented by symbols with error bars, as indicated in the legend, while the simulation results are depicted as red open triangles. It is observed that PACIAE 4.0 provides a satisfactory reproduction of the pseudorapidity density distributions of charged particles, $dN_{ch}/d\eta$, for NSD events at the three collision energies. Notably, the model provides an excellent reproduction of the $dN_{ch}/d\eta$ distribution in $p\bar{p}$ collisions at 0.54 TeV within the pseudorapidity region ($|\eta| < 2$), while the experimental data falls off slightly faster than the simulation results in the forward and backward regions. By comparing with the experimental data at $\sqrt{s}=$0.63 TeV, we conjecture that the $dN_{ch}/d\eta$ measurements at 0.54 TeV in the forward/backward regions could have larger uncertainties, although this is not conclusive. The pseudorapidity density distributions from CDF and P238 collaborations agree with each other within errors, however their central values do not overlap in the forward/backwork regions. As in $pp$ collisions, the pseudorapidity density distributions of charged particles in NSD $p\bar{p}$ collisions exhibit only a weak $\eta$-dependence at midrapidity. 

Figure~\ref{fig2-1} shows the transverse momentum spectra of charged particles produced in NSD $p\bar{p}$ collisions at $\sqrt{s}=$ 0.54, 0.63, and 1.8 TeV. The experimental data from the UA1 and CDF collaborations are represented by black solid triangles (0.54 TeV), squares (0.63 TeV), and circles (1.8 TeV) with error bars. The corresponding simulation results are depicted by red open triangles, squares, and circles, respectively. It can be seen that PACIAE 4.0 reproduces the transverse momentum spectra of charged particles in a satisfactory manner, capturing the general trend and magnitude across different collision energies. The maximum deviation from experimental data is 50\%. In the region $p_{\rm T}<2$ GeV at $\sqrt{s}=$ 0.54 and 1.8 TeV, the simulated results are consistent with the experimental data. However, the simulated results at $\sqrt{s}=$0.63 TeV are slightly larger than the experimental data in low $p_T$ region. We have investigated this discrepancy. The pseudorapidity density distributions of charged particles shown in Fig. \ref{fig1} are derived by integrating the transverse momentum spectra over the full $p_{\rm T}$ range. Based on the results for $\sqrt{s}=0.63$ TeV in Figs. \ref{fig1} and \ref{fig2-1}, we conclude that the experimental data exceeds the simulation results in the very low $p_T$ region.

Having validated the assumption that the model parameters previously determined from $pp$ collisions are applicable to the same event types in $p\bar{p}$ collisions, we now investigate the effects of matter-antimatter interactions on particle production. We simulate INEL $pp$ and $p\bar{p}$ collisions at given energies using PACIAE 4.0. The INEL event class is chosen as it includes single-diffractive events where only one $p(\bar{p})$ is shattered. It probably leads the effects of matter-antimatter interactions more pronounced for the baryon production. We compare the transverse momentum spectra of identified particles ($\pi^+$, $K^+$, protons, and neutrons) at midrapidity $|\eta|<0.5$ from $p\bar{p}$ and $pp$ collisions at $\sqrt{s}=$ 0.1, 0.2, and 2.76 TeV, as shown in Fig.~\ref{fig2}. The spectra from $p\bar{p}$ collisions are represented by blue solid symbols (triangles for 0.1 TeV, circles for 0.2 TeV, and squares for 2.76 TeV), while the results from $pp$ collisions are shown by red open symbols. For better visualization, the spectra are scaled by the factors indicated in the legend, respectively. It is observed that the transverse momentum spectra of $\pi^+$ and $K^+$ are nearly identical in $pp$ and $p\bar{p}$ collisions at all energies. Similary, at LHC energy ($\sqrt{s}=2.76$ TeV), the spectra of protons and neutrons from the two collision systems are indistinguishable. However, as the collision energy decreases, the transverse momentum spectra of protons and neutrons from $pp$ collisions become higher than those from $p\bar{p}$ collisions, with the discrepancy growing as the collision energy decreases.  

The observed results can generally be understood within the framework of the quark model. Pions and kaons ($\pi^+$, $K^+$) are the lightest mesons, composed of a quark-antiquark pair, and constitute the majority of produced particles. Their constituent quarks are predominantly excited from the vaccum during collisions. Consequently, their production depends more on the available collision energy than on the nature of the initial-state nucleons, leading to nearly identical productions in both $pp$ and $p\bar{p}$ collisions at all energies considered. In contrast, protons and neutrons are baryons, and their production is linked to both the collision energy and the initial-state nucleons. At very high energies such as those at the LHC, an abundance of quarks is excited from the vaccum and the contribution from the valence quarks of the initial nucleons becomes negligible. However, as the collision energy decreases, the role of valence quarks becomes more significant. A $pp$ collision system contains more valence quarks ($u$ and $d$) than a $p\bar{p}$ system. Therefore, the production of $p$($n$) is enhanced in $pp$ collisions compared to $p\bar{p}$ collisions, respectively. This valence quark effect becomes more pronounced as the collision energy decreases.

To further investigate the influence of valence quarks and initial-state nucleons on the production of protons and neutrons, we compare results from INEL and NSD events in both $pp$ and $p\bar{p}$ collisions. The comparison is motivated by the potential for single-diffractive events to affect the production of protons from the initial nucleons in the INEL event class as discussed before. Figure \ref{fig3} shows the transverse momentum spectra of protons and neutrons from $pp$ and $p\bar{p}$ collisions at $\sqrt{s}=0.1$ TeV. Open and solid symbols represent protons and neutrons, respectively, as detailed in the legend. It is observed that in $pp$ collisions, the transverse momentum spectra of protons and neutrons are nearly identical, with protons being slightly higher than neutrons for both INEL and NSD event classes. This observation can be understood within the quark model: the number of valence $u$ quarks is larger than that of valence $d$ quarks, making proton production more probable than neutron production. This result is also consistent with charge conservation, as the initial charge in a $pp$ collision is $+2$. A similar explanation can also apply to the difference between proton and neutron yields in INEL and NSD $p\bar{p}$ collisions. Because the relative difference between protons and neutrons is similar in both INEL and NSD events. This similarity cannot be explained by the single diffractive process or the net-zero valence quarks in the initial state.
Furthermore, the transverse momentum spectra of protons (neutrons) from NSD events are systematically higher than those from INEL events. This can be attributed to the nature of the event classes: NSD events represent more central, ``fierce'' collisions than INEL events, which include peripheral diffractive processes. Consequently, more quarks are excited from the vacuum in NSD events, leading to greater particle production, as demonstrated by pseudorapidity density distributions in $pp$ collisions \cite{Xie:2025vnh, Xie:2025zoi} and references therein. Finally, the difference in nucleon spectra from NSD (INEL) events is larger in $p\bar{p}$ collisions than in $pp$ collisions. This could be associated with the initial-state baryon number: a net baryon number conservation constraint exists in $pp$ collisions, whereas no such net constraint exists in $p\bar{p}$ collisions.

\section{CONCLUSIONS}\label{con}
The PACIAE 4.0 model has been employed to systematically simulate pseudorapidity density distributions and transverse momentum spectra of charged particles in NSD $p\bar{p}$ collisions at $\sqrt{s} = 0.54$, $0.63$, and $1.8$ TeV. Crucially, the model parameters were taken directly from previous $pp$ studies with same event type, with no additional tunning for the $p\bar{p}$ system. The model successfully reproduces the experimental data from the UA1, CDF, and P238 collaborations. We further investigate matter-antimatter interactions by comparing the transverse momentum spectra of identified particles in INEL $p\bar{p}$ and $pp$ collisions at $\sqrt{s} = 0.1$, $0.2$, and $2.76$ TeV, as well as in NSD collisions at $\sqrt{s} = 0.1$ TeV. The spectra of $\pi^+$ and $K^+$ mesons are nearly identical in $pp$ and $p\bar{p}$ collisions at all energies. In contrast, the production of protons and neutrons in $pp$ collisions exceeds that in $p\bar{p}$ collisions, with the discrepancy increasing at lower energies. Furthermore, nucleon yields are higher in NSD events than in INEL events. All these findings are well explained within the framework of the quark model. In conclusion, this work demostrates that the PACIAE 4.0 model is a versatile and reliable tool for studying high-energy collision physics.

\section*{Acknowledgement}
 This work is supported by the National Natural Science Foundation of China under grant Nos. 11447024, 11505108 and 12375135, and by the 111 project of the foreign expert bureau of China. Y.L.Y. acknowledges the financial support from Key Laboratory of Quark and Lepton Physics in Central China Normal University under grant No. QLPL201805 and the Continuous Basic Scientific Research Project (No, WDJC-2019-13). W.C.Z. is supported by the Natural Science Basic Research Plan in Shaanxi Province of China (No. 2023-JCYB-012). H.Z. acknowledges the financial support from Key Laboratory of Quark and Lepton Physics in Central China Normal University under grant No. QLPL2024P01.

\bibliography{reference.bib}

\end{document}